\documentclass[twocolumn]{aastex631}
\usepackage{multirow}
\submitjournal{ApJ}

\shorttitle{Galaxy Morphology in CEERS at $z<$3}
\shortauthors{Yao et al.}

\begin{document}

\title{Evolution of Non-parametric Morphology of Galaxies in the JWST CEERS Field at $z\simeq$0.8-3.0}

\correspondingauthor{Xu Kong}
\email{yaoyao97@mail.ustc.edu.cn, xkong@ustc.edu.cn}

\author[0000-0002-6873-8779]{Yao Yao}
\affiliation{Deep Space Exploration Laboratory / Department of Astronomy, University of Science and Technology of China, Hefei 230026, China}

\author{Jie Song}
\affiliation{Deep Space Exploration Laboratory / Department of Astronomy, University of Science and Technology of China, Hefei 230026, China}

\author{Xu Kong}
\affiliation{Deep Space Exploration Laboratory / Department of Astronomy, University of Science and Technology of China, Hefei 230026, China}
\affiliation{School of Astronomy and Space Sciences, University of Science and Technology of China, Hefei, 230026, China}

\author{Guanwen Fang}
\affiliation{Institute of Astronomy and Astrophysics, Anqing Normal University, Anqing, 246133, China}

\author{Hong-Xin Zhang}
\affiliation{Deep Space Exploration Laboratory / Department of Astronomy, University of Science and Technology of China, Hefei 230026, China}

\author{Xinkai Chen}
\affiliation{Deep Space Exploration Laboratory / Department of Astronomy, University of Science and Technology of China, Hefei 230026, China}
\affiliation{Kavli Institute for Astronomy and Astrophysics, Peking University, Beijing, 100871, China}



\begin{abstract}

Galaxy morphology is one of the most fundamental ways to describe galaxy properties, but the morphology we observe may be affected by wavelength and spatial resolution, which may introduce systematic bias when comparing galaxies at different redshift. Taking advantage of the broad wavelength coverage from optical to near-IR and high resolution NIRCam instrument of JWST, we measure the non-parametric morphological parameters of a total of 1376 galaxies at $z\simeq$0.8-3.0 in the CEERS field through an optimized code called {\tt\string statmorph\_csst}. We divide our sample into three redshift intervals and investigate the wavelength- and redshift-dependence of the morphological parameters. We also explore how the widely-used galaxy type classification methods based on the morphological parameters depend on wavelength and spatial resolution. We find that there are variations in all morphological parameters with rest-frame wavelength ($\lambda_{\rm rf}$), especially at the short wavelength end, and the $\lambda_{\rm rf}$ mainly affects the classification between late-type and early-type galaxy. As the $\lambda_{\rm rf}$ increases, the galaxies on the $G-M_{20}$ diagram move to the upper left with a slope of -0.23$\pm$0.03 on average. We find that spatial resolution mainly affects the merger identification. The merger fraction in F200W resolution can be $\ga$2 times larger than that in F444W resolution. Furthermore, We compare the morphological parameter evolution of galaxies with different stellar masses. We find that there are differences in the morphological evolution of high- and low-mass (log$M_*\geqslant$10 and 9$<$log$M_*<$10) galaxies in the studied redshift range, which may be caused by their different evolution paths. 

\end{abstract}



\section{Introduction} \label{sec:intro}

Morphology of galaxies is a fundamental aspect that provides crucial insights into the structure and formation of galaxies. It allows a direct representation of the spatial distribution of stars and dust in galaxies, which holds the key to understanding  their evolution and assembly history. The galaxy morphology of the local universe has rich diversity. \cite{Hubble1926} has classified galaxies into the now well-known tuning fork through their visual structures. His classification includes three distinct groups: ellipticals, spirals, and irregulars, where ellipticals are also usually called Early-Type Galaxies (ETGs), and spirals and irregulars are also usually called Late-Type Galaxies (LTGs). There are not only significant differences in morphology between ETGs and LTGs but also significant differences in star formation rate and other physical properties.

In the last decades, thanks to the sky surveys of Hubble Space Telescope (HST) and ground-based telescopes, the relation between galaxy morphology and galaxy physical evolution have been researched deeper \citep[e.g.,][]{Kauffmann2003,Baldry2004,Conselice2006,Mortlock2013} and more methods to classify galaxy morphology have been developed, including S\'{e}rsic model \citep{Sersic1963}, bulge/disk decomposition \citep{Freeman1970,Kent1985}, and non-parametric measurements \citep[e.g., ][]{Conselice2003,Lotz2004,Freeman2013,Ferrari2015}. These methods are more objective than manual visual inspection and enable comparison between results from different researchers. In recent years, with the rapid evolution of computer technology, machine-learning algorithms for galaxy morphological classification and merger identification have been widely developed and utilized. \citep[e.g., ][]{DominguezSanchez2018,Tang2020,Yao2022a,Fang2023}. At the same time, the powerful computing power has also facilitated a lot of galaxy morphological studies based on cosmological simulation \citep[e.g.,][]{Snyder2015,Bignone2017,RodriguezGomez2019,Bignone2020}, which provide a direct way to link observed morphology with physical processes.

However, limited by the spatial resolution, sensitivity, and red wavelength range of HST (reddest filter F160W), the morphological study of high-redshift galaxies in rest-frame optical band stops at $z\simeq$3. The new James Webb Space Telescope (JWST) solves these problems well. Its superior spatial resolution and longer wavelength filter set (reddest filter F444W of NIRCam) not only allow us to explore galaxy structure at $z>$6 \citep[e.g., ][]{Ferreira2022,Kartaltepe2023,Treu2023} but also make the images of galaxies at lower redshifts have higher quality and wider wavelength range, which can reduce the measurement uncertainties caused by the resolution, signal-to-noise ratio (S/N) and morphological K-correction.

In this work, we make use of the JWST's observation data and study the non-parametric morphology of galaxies at $z\simeq$0.8-3.0. This paper is organized as follows. In Section \ref{sec:data}, we describe our selection criteria of the sample and the non-parametric measurements from the data. In Section \ref{sec:result}, we present our results of non-parametric morphology measurement and discuss the effects of the rest-frame wavelength ($\lambda_{\rm rf}$) and the point spread function (PSF) on morphology and the differences between galaxies with different stellar masses ($M_*$) at different redshifts. In Section \ref{sec:conclusion}, we summarize the main results of this work.
Throughout this paper, we adopt a $\Lambda$-CDM cosmology with $\Omega_\Lambda$= 0.7, $\Omega_m$=0.3, and H$_0$=70 km s$^{-1}$ Mpc$^{-1}$.

\section{Data and Measurements} \label{sec:data}
\subsection{CEERS Images and Catalogs}
CEERS\footnote{\url{https://ceers.github.io/index.html}} (Finkelstein et al., in prep.) is one of 13 early release science (ERS) surveys of JWST. It is designed to obtain data covering the EGS \citep[Extended Groth Strip, ][]{Davis2007} extragalactic deep field early in Cycle 1. The observations, tests, and data products will pave the way for Cycle 2 observations.
CEERS is based around a mosaic of 10 NIRCam \citep{Rieke2005} pointings, with six obtaining NIRSpec \citep{Jakobsen2022} in parallel and four with MIRI \citep{Rieke2015} in parallel. The NIRCam images of CEERS includes 7 bands, from the short wavelength (SW) detector (F115W, F150W, and F200W) and the long wavelength (LW) detector (F277W, F356W, F410M, and F444W).

Here we make use of the first four CEERS NIRCam pointings, obtained on 2022 June 21, known as CEERS1, CEERS2, CEERS3, and CEERS6. We directly use the final mosaic images and catalogs provided by \cite{Valentino2023}\footnote{\url{https://s3.amazonaws.com/grizli-v2/JwstMosaics/v4/index.html\#ceers}}, which is processed with the {\tt\string grizli} \citep{Brammer2022} software pipeline. Besides zeropoint correction\footnote{\url{https://github.com/gbrammer/grizli/pull/107}} relative to {\tt\string jwst\_0942.pmap}, the 1/f, snowballs, and wisp correction have also been applied by {\tt\string grizli}.
The pipeline also uses {\tt\string SEP} \citep{Barbary2016}, a Python library implementing {\tt\string Source-Extractor} \citep{Bertin1996} to extract sources from the detection image combining of all the ``wide'' NIRCam LW filters available (typically F277W+F356W+F444W) optimally weighted by their noise maps. It also outputs JWST+HST photometric catalogs and {\tt\string EAZY-py}\footnote{\url{https://github.com/gbrammer/eazy-py}} \citep{Brammer2008} photometric redshift catalogs for all pointings. Spectroscopic redshifts have also been added to these catalogs, if available. The sampling rate of the final SW mosaic images is 0\arcsec.02 pixel$^{-1}$, and that of LW images is 0\arcsec.04 pixel$^{-1}$.

\subsection{Sample Selection} \label{subsec:sample}

We select a sample to analyze morphology from the photometric and redshift catalogs of \cite{Valentino2023}, which contain 26438 sources. We select galaxies that have {\tt\string mag\_auto}$<$27.0 in detection image, and cover at least one pixel on the segmentation maps. A total of 2969 sources meet the above conditions, of which 75 sources do not have effective photometric or spectroscopic redshift and $M_*$ (in unit of $M_\odot$) in the above catalogs. We divided the remaining 2894 sources into 3 redshift bins: 0.8-1.3, 1.3-2.0, and 2.0-3.0. The reason why we choose these values is that we can match the NIRCam filters of different bins at a similar $\lambda_{\rm rf}$. In addition, the reason why we do not choose a higher redshift is that the sample size at the higher redshift is limited, the strong cosmic dimming effect, and few filters can be matched with the low-redshift galaxies. Thus, in these three redshift bins, we have four bands that can be matched at a similar $\lambda_{\rm rf}$, at $\sim$570 nm (close to $V$), $\sim$760 nm (close to $i$), $\sim$1000 nm, and $\sim$1340 nm respectively.

We use the catalog containing 2894 sources as the input of our process pipeline for morphological parameter measurement (see Section \ref{subsec:measurement}). According to the result of our measurement, we exclude sources that have an invalid morphological parameter in any filters, mainly contributed by the diffraction spikes of starlight. \cite{Lotz2004,Lotz2006} and \cite{Treu2023} have pointed out that the morphological parameter measurement accuracy is affected when the average S/N per pixel ($\langle$S/N$\rangle$) goes below 2. So we only select sources whose $\langle$S/N$\rangle$ are higher than 2 in all bands we analyzed. As a result, the number of sources decreased from 2894 to 2068. Section 4.3.2 of \cite{RodriguezGomez2019} describes the method of obtaining $\langle$S/N$\rangle$.

\begin{figure}[ht!]
    \centering
    \includegraphics[width=1\columnwidth]{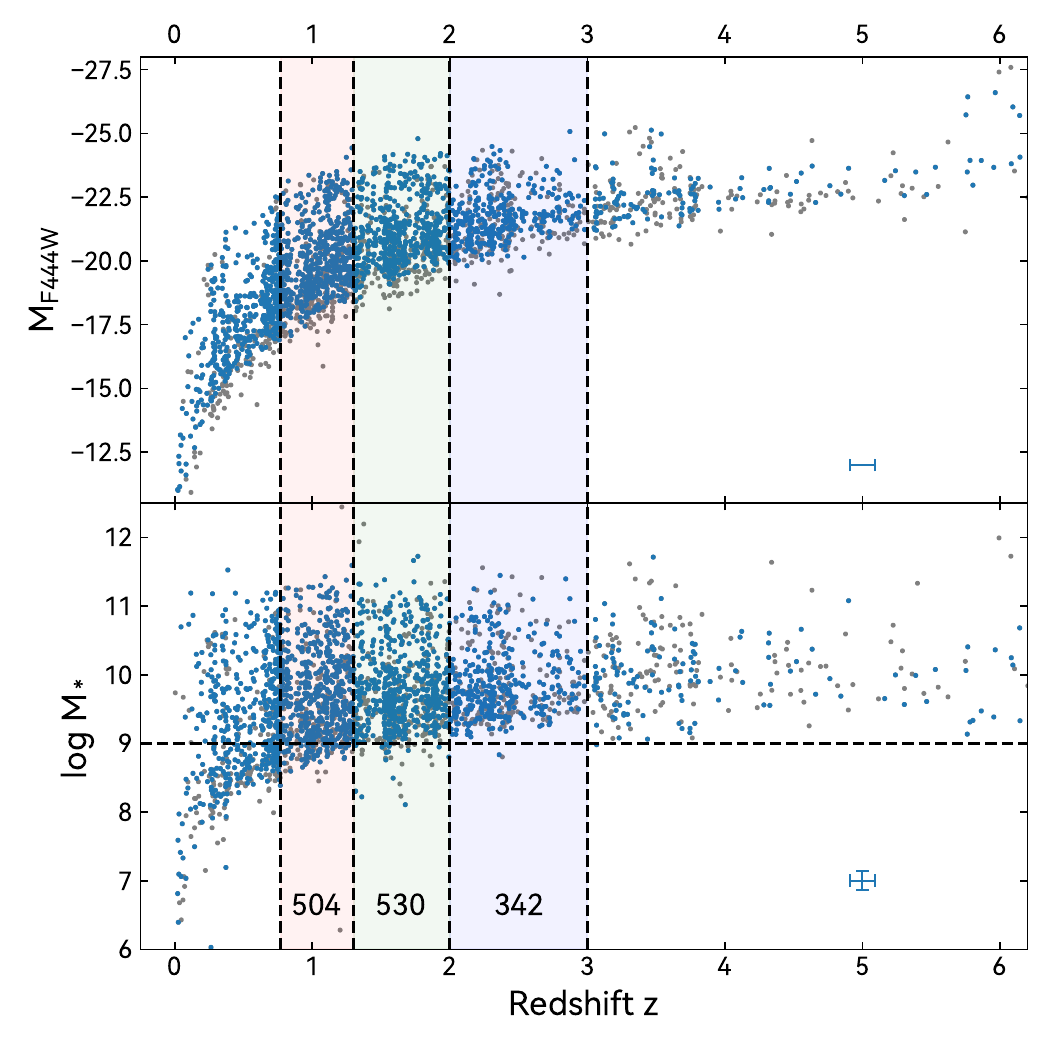}
    \caption{The distribution of the M$_{\rm F444W}$ and the $M_*$ of our galaxies with redshift. The blue dots represent sources with high $\langle$S/N$\rangle$s and valid morphological parameters (2068). The gray dots represent other sources removed due to not meeting the conditions above (826). The horizontal dashed black line represents our cut of $M_*$. The three different colored rectangular regions divided by vertical dashed black lines represent our three redshift bins, in which the numbers represent the number of galaxies.
    \label{fig:z_mass}}
\end{figure}

Figure \ref{fig:z_mass} shows the distribution of the absolute magnitude of F444W(M$_{\rm F444W}$) and the $M_*$ of our galaxies with redshift, and Table \ref{tab:bins_summary} summaries the details of three redshift bins which we select for our main research. We only analyze galaxies whose log$M_*>$9 to balance the completeness and range of $M_*$, ensuring that the sample size is greater than 1000 and the lower redshift bin is mass-completed. Finally, there are 504, 530, and 342 galaxies in each bin, resulting in 1376 galaxies in total for following analysis, in which 299 galaxies have spectroscopic redshifts and the remaining 1077 galaxies have a mean uncertainty of photometric redshift of $\sim$0.097. The mean uncertainty of log$M_*$ of all galaxies in our three bins is $\sim$0.145 and the mean uncertainty of M$_{\rm F444W}$ is $\sim$0.007.

\begin{deluxetable*}{ccccccccc}
	\tablecaption{Summary of redshift bins\label{tab:bins_summary}}
	\tablewidth{0pt}
	\tablehead{
		\colhead{Redshift bin} & \colhead{Mean redshift} & \colhead{Total} & \colhead{Filter} &\colhead{Resolution} & \colhead{$\lambda_{\rm rf}$} & \multicolumn{3}{c}{$G-M_{20}$ classification} \\
        \colhead{$z$} & \colhead{$\left\langle z \right\rangle$} & \colhead{} & \colhead{} & \colhead{} & \colhead{(nm)} & \colhead{Merger} & \colhead{LTG} & \colhead{ETG}
	}
	\decimalcolnumbers
	\startdata
	         &      &     & F115W & F200W & $\sim$570 & 72 (14.3\%) & 336 (66.7\%) & 96 (19.0\%) \\ 
              &      &     & F115W & F444W & $\sim$570 & 33 (6.5\%) & 426 (84.5\%) & 45 (8.9\%) \\
    0.8-1.3 & 1.05 & 504 & F150W & F444W & $\sim$760 & 25 (5.0\%) & 435 (86.3\%) & 44 (8.7\%) \\
              &      &     & F200W & F444W & $\sim$1000 & 23 (4.6\%) & 436 (86.5\%) & 45 (8.9\%) \\
              &      &     & F277W & F444W & $\sim$1340 & 22 (4.4\%) & 428 (84.9\%) & 54 (10.7\%) \\
    \hline
              &      &     & F150W & F200W & $\sim$570 & 111 (20.9\%) & 329 (62.1\%) & 90 (17.0\%) \\ 
              &      &     & F150W & F444W & $\sim$570 & 36 (6.8\%) & 460 (86.8\%) & 34 (6.4\%) \\
    1.3-2.0 & 1.65 & 530 & F200W & F444W & $\sim$760 & 26 (4.9\%) & 469 (88.5\%) & 35 (6.6\%) \\
              &      &     & F277W & F444W & $\sim$1000 & 31 (5.8\%) & 443 (83.6\%) & 56 (10.6\%) \\
              &      &     & F356W & F444W & $\sim$1340 & 37 (7.0\%) & 438 (82.6\%) & 55 (10.4\%) \\
    \hline
              &      &     & F200W & F200W & $\sim$570 & 92 (26.9\%) & 201 (58.8\%) & 49 (14.3\%) \\ 
              &      &     & F200W & F444W & $\sim$570 & 19 (5.6\%) & 309 (90.4\%) & 14 (4.1\%) \\
    2.0-3.0 & 2.36 & 342 & F277W & F444W & $\sim$760 & 17 (5.0\%) & 307 (89.8\%) & 18 (5.3\%) \\
              &      &     & F356W & F444W & $\sim$1000 & 37 (7.0\%) & 438 (82.6\%) & 55 (10.4\%) \\
              &      &     & F444W & F444W & $\sim$1340 & 28 (8.2\%) & 293 (85.7\%) & 21 (6.1\%) \\
	\enddata
	\tablecomments{From left to right, the columns (1), (2), and (3) correspond to the redshift range of each bin, mean redshift of all galaxies in the bin, and the quantity of galaxy with log$M_*>9$. The columns (4), (5), and (6) correspond to the NIRCam filter used for the measurement, the target filter used to match PSF, and the rest-frame pivot wavelength (approximation) of the filter at the mean redshift. The columns (7)-(9) are the quantity and fraction of different types of galaxy classified by the $G-M_{20}$ diagram.}
\end{deluxetable*}

\subsection{Non-parametric Morphology Measurement} \label{subsec:measurement}
Non-parametric morphology measurement is designed to capture the major features of the underlying structures of galaxies, without any assumed light distribution model (e.g. S\'{e}rsic profile). At high redshift universe, galaxies are usually irregular and cannot be fitted by a simple model \citep{Conselice2014}. Non-parametric measurement is ideal for characterizing galaxy evolution over many epochs and has been applied in many high-redshift works \citep[e.g., ][]{Law2012,Ferreira2022a,Treu2023}.

We analyze all wide bands of the CEERS NIRCam images (F115W, F150W, F200W, F277W, F356W, and F444W) by the {\tt\string statmorph} code \citep{RodriguezGomez2019}, which calculates non-parametric morphological diagnostics such as the Concentration-Asymmetry-Smoothness system \citep[CAS, ][]{Conselice2003} and Gini-$M_{20}$ system \citep[$G$–$M_{20}$, ][]{Lotz2004}. We have modified the code of {\tt\string statmorph} to improve performance, resulting in a 3-5 times improvement in computation speed without sacrificing measurement accuracy. This optimization was mainly accomplished through acceleration by {\tt\string Cython} \citep{Behnel2011}. Appendix \ref{sec:performance} shows our optimization performance and a comparison between the outputs of the original version of  {\tt\string statmorph} and our optimized code. In addition, the modified version also includes the call to {\tt\string Source-Extractor}. Users can enter parameters of {\tt\string Source-Extractor} to do photometry and generate a segmentation map, or skip this step and provide their own segmentation map. Furthermore, we also add some extended optional morphological parameters (e.g., $\xi$, \citealt{Papovich2003}; $\Psi$, \citealt{Law2007}; $G_2$, \citealt{Rosa2018}), which can be selectively calculated according to the user's specific needs. Finally, we change the flag used for warnings to a bitmap format, which can store different types of warnings rather than just indicating their presence or absence. All the above optimizations and improvements are made to meet the requirements of the China Space Station Telescope \citep[CSST, ][]{Zhan2011,Gong2019} survey in the future, so we rename the modified code to {\tt\string statmorph\_csst}\footnote{Optimized code is available at \url{https://github.com/pentyum/statmorph_csst} and \url{https://gitee.com/pentyum/statmorph_csst}}. CSST is a 2-meter space telescope planned to be launched in the early 2020s. It will operate in the same orbit as the China Manned Space Station and survey 17,500 deg$^2$ sky area over 10 years of operation. In this work, we directly use the segmentation map provided by {\tt\string grizli} as the input of {\tt\string statmorph\_csst}.

\subsubsection{CAS Statistics}
The most-widely used non-parametric morphological indicator is the CAS statistics. It includes three parameters, concentration ($C$), asymmetry ($A$), and smoothness ($S$, also clumpiness). The definitions and the corrections of these parameters have changed and developed over the past decades \citep[e.g., ][]{Kent1985,Abraham1996,Bershady2000,Conselice2003,Conselice2003a,Lotz2004,Conselice2008,Law2012}, and the definition given in \cite{Conselice2014} is generally adopted in recent literature. 

\subsubsection{$G-M_{20}$ Statistics}
Another set of widely used non-parametric morphological indicators is the $G-M_{20}$ classification system \citep{Abraham2003,Lotz2004,Lotz2008}. The measurement of these parameters does not need to provide a predefined location of the galactic center, which is particularly desirable for irregular and merging galaxies at high redshift.

To calculate the $G-M_{20}$ statistics, {\tt\string statmorph} creates a new $G-M_{20}$ segmentation map that only selects pixels whose brightness are brighter than the average brightness at major-axis radius of $r_p$. This procedure is the same as \cite{Lotz2004}, and its detailed implements by {\tt\string statmorph} are described in Section 4.4.1 of \cite{RodriguezGomez2019}.





\subsection{PSF Matching} \label{subsec:psf_matching}
To accurately measure the morphological parameters of galaxies in each NIRCam filter, it is essential to account for the difference of PSFs across filters. Since JWST is a space telescope, its PSF is diffraction-limited PSF. The longer the wavelength is, the larger the FWHM of the PSF is. According to the simulation by {\tt\string WebbPSF} \citep{Perrin2014}, FWHMs of the NIRCam bands vary from 0\arcsec.037 (F115W) to 0\arcsec.140 (F444W), which may introduce systematic wavelength-dependent bias in the measurement of morphological parameters \citep[e.g., ][]{Bershady2000,Lotz2004,Law2012}, especially in $A$ \citep{Bignone2020}. Therefore, it is necessary to perform PSF matching, which creates convolution kernels from PSFs with small FWHM to PSFs with large FWHM. The following in this subsection describes our PSF matching procedure. The overall process is similar to \cite{Yao2022}, but some optimizations have been made.

The PSFs we use are from {\tt\string grizli-psf-library}\footnote{\url{https://github.com/gbrammer/grizli-psf-library/tree/main/ceers}}. These PSFs are sampled at a pixel scale of 0\arcsec.04.
When we apply these PSFs to the SW images of NIRCam, whose pixel scale is 0\arcsec.02. We resample the PSFs to match the pixel scale.

We make use of the {\tt\string photutils}\footnote{\url{https://github.com/astropy/photutils}} \citep{Bradley2022} package to create a convolve kernel from short-wavelength image to long-wavelength. We convolve all images with wavelength shorter than F444W to the F444W, whose PSF is the largest among all filters. In order to study the influence of different PSFs, besides the convolution to F444W, we also convolve the F115W and F150W image to the F200W whose PSF is the largest among the SW detectors.

We use an optimized method to create convolve kernels. \cite{Aniano2011} provide two parameters to measure kernel performance. One is $D$, describing the accuracy in redistribution of PSF power. The other is $W_-$, describing the negative values of the kernel. The lower the two values, the better. For a good PSF matching performance, $D\textless$0.1 and $W_-\textless$1 are required. We take this as our optimization goal and build a simple loss function with $D$ and $W_-$ as parameters. We find the parameter of each window function\footnote{\url{https://photutils.readthedocs.io/en/stable/psf_matching.html}} provided by {\tt\string photutils} that minimizes the loss function, and then compare the minimum value of the loss function generated by different types of window function, and then select the lowest one. Table \ref{tab:kernel_performance} lists the types of window functions we finally adopt and the corresponding $D$ and $W_-$ during the PSF matching process.

\begin{deluxetable}{ccccc}
	\tablecaption{Window type and the kernel performance of PSF matching\label{tab:kernel_performance}}
	\tablewidth{0pt}
	\tablehead{
		\colhead{Original filter} & \colhead{Target filter} & \colhead{Window type} & \colhead{$D$} & \colhead{$W_-$}
	}
	\decimalcolnumbers
	\startdata
	F115W & F444W & Tukey & 0.032 & 0.069 \\
    F150W & F444W & CosineBell & 0.047 & 0.084 \\
    F200W & F444W & CosineBell & 0.074 & 0.068 \\
    F277W & F444W & TopHat & 0.017 & 0.345 \\
    F356W & F444W & TopHat & 0.019 & 0.629 \\
    \hline
    F115W & F200W & CosineBell & 0.080 & 0.341 \\
    F150W & F200W & CosineBell & 0.088 & 0.386 \\
	\enddata
	\tablecomments{Photutils provides the following window function types: HanningWindow, TukeyWindow, CosineBellWindow, SplitCosineBellWindow, and TopHatWindow. Here we have tried all the window function type, except the SplitCosineBellWindow, because it contains multiple parameters, which is not convenient to use {\tt\string scipy.optimize.minimize\_scalar}, and other types have been enough to meet our optimization goals.}
\end{deluxetable}

Then we use the {\tt\string reproject}\footnote{\url{https://github.com/astropy/reproject}} package to project the convolved SW images (0\arcsec.02 pixel$^{-1}$) to the same sampling rate as the LW images (0\arcsec.04 pixel$^{-1}$). In this way, we can get multi-band images with the same spatial resolution and sampling rate.
 
Given the redshift range ($z$=0.8-3.0) of the galaxies measured by us, the physical scale of given angular size does not vary much (7.5-8.5 kpc arcsec$^{-1}$), so we do not attempt to make further redshift-dependent adjustment of the PSF. The physical scale corresponding to the FWHM of the PSF in F444W band is 1.05-1.19 kpc, and the physical scale of the pixel in our final images is 0.30-0.34 kpc.

\section{Results and Discussion} \label{sec:result}

\subsection{Morphology Evolution with Rest-frame Wavelength} \label{subsec:wavelength_evolution}

It is well known that the color of different age of stellar population is different. At shorter wavelengths (e.g. $U$-band), the morphology and quantitative structures trace the distribution of younger stellar populations, whereas longer wavelengths (e.g. $R$-band) tend to trace older stellar populations that usually dominate the total stellar mass.  Furthermore, the distribution of dust and emission nebula can also affect the morphology at different $\lambda_{\rm rf}$ \citep{Baes2020,Nersesian2023}. Therefore, to ensure a fair comparison of galaxy morphologies at different redshift, it is necessary to make a ``K-correction'' \citep{Kuchinski2001,Conselice2003} when analyzing them. \cite{Wuyts2012} have analyzed some morphological parameters ($R_e$, $C$, $G$, and, $M_{20}$) of massive SFGs in the HST GOODS-South field ($z$=0.5-2.5), and finds that SFGs to be more concentrated, smaller, and smoother in red bands or their stellar mass distribution than in blue bands. These differences can be explained by the inside-out growth scenario in combination with a "Christmas tree" model \citep{Lowenthal1997}.

\begin{figure*}[ht!]
    \centering
    \includegraphics[width=1\textwidth]{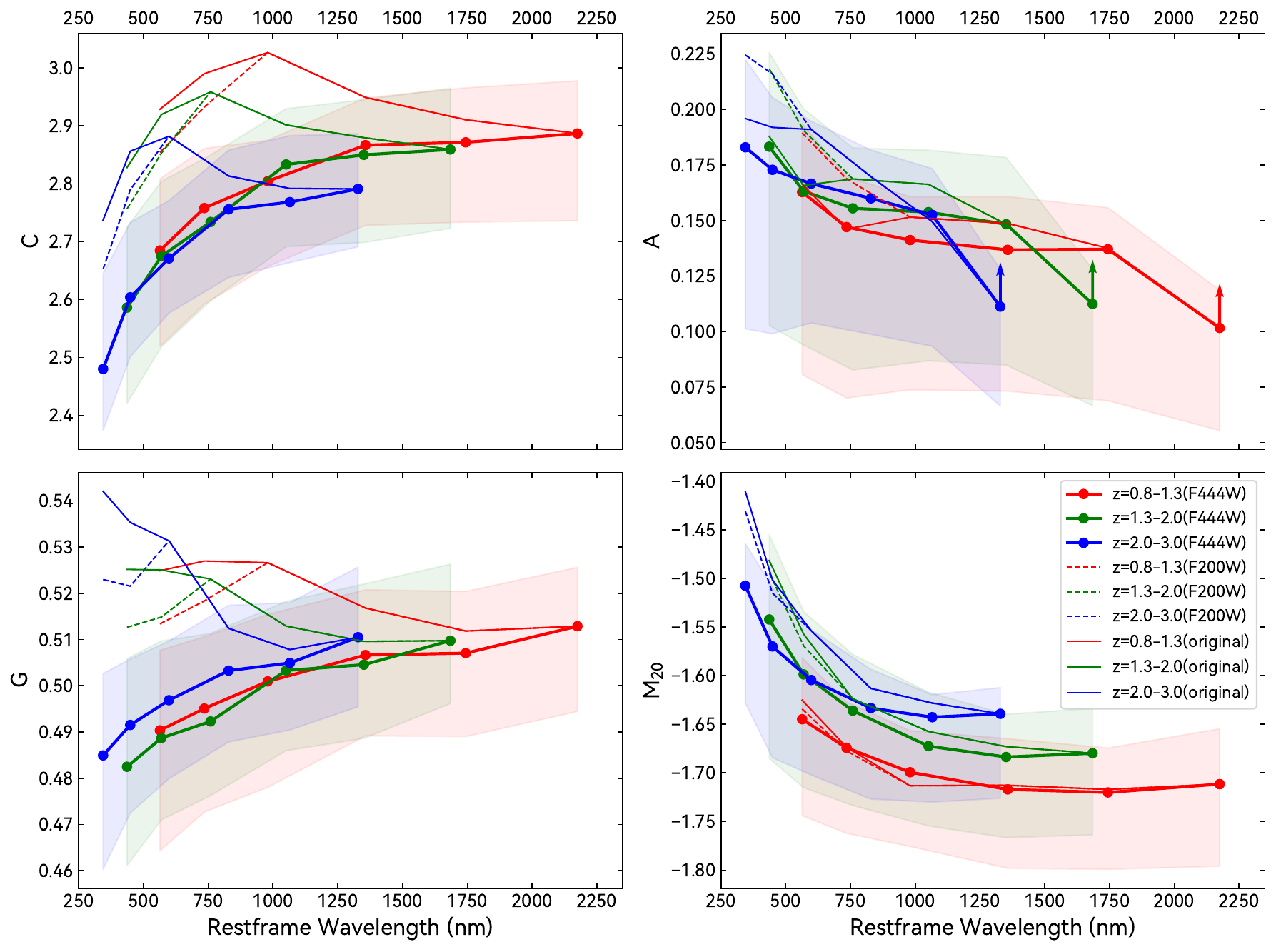}
    \caption{The four morphological parameters, $C$, $A$, $G$, and $M_{20}$ vary with $\lambda_{\rm rf}$ in the three redshift bins colored by red, green, and blue individually. The lines and dots represent the mean value of all galaxies in each bin. The semi-transparent filled areas indicate their 32\% and 68\% (1$\sigma$) quantiles. The thick solid lines represent the values that have been matched to F444W PSF. The thin dashed lines represent the values that have been matched to F200W PSF, while the thin solid lines represent values without any PSF matching. The arrows at long wavelength ends are the rough corrections of the underestimated $A$ in F444W.
    \label{fig:wavelength_evolution}}
\end{figure*}

Thanks to the wide NIR wavelength range of JWST instruments, we are able to investigate the morphology of galaxies and study their evolution at different redshifts at similar $\lambda_{\rm rf}$. Figure \ref{fig:wavelength_evolution} illustrates how the average of the four morphological parameters, $C$, $A$, $G$, and $M_{20}$, vary with $\lambda_{\rm rf}$. Our analysis reveals that all morphological parameters of all galaxies, regardless of their redshift, are influenced by the $\lambda_{\rm rf}$. Specifically, the morphological parameters have a stronger wavelength dependence at shorter wavelengths ($<$1000nm), while at longer wavelengths ($>$1000nm), the parameters become nearly independent of the $\lambda_{\rm rf}$.

At the same resolution, we observe that the $C$ values noticeably increase as the wavelength increases. Furthermore, the $C$ values exhibit little difference among galaxies at different redshifts at short $\lambda_{\rm rf}$. However, at longer $\lambda_{\rm rf}$ ($\sim$1340nm), the $C$ values of galaxies at $z$=2.0-3.0 are 0.076 lower than those of galaxies at $z$=0.8-1.3. For each redshift bin we examined ($z$=0.8-1.3, $z$=1.3-2.0, $z$=2.0-3.0), the $C$ values increase by $\sim$0.18 ($\sim$6.8\%), $\sim$0.18 ($\sim$6.5\%), and $\sim$0.12 ($\sim$4.5\%), respectively, from rest-frame $\sim$570nm to $\sim$1340nm. \cite{Wuyts2012} have also found that the wavelength dependence of $C$ at high redshift is weaker than low redshift, but the amplitude of the change in our work is lower than that obtained by \cite{Wuyts2012}. This may be because we calculate the difference between the optical and NIR bands, while \cite{Wuyts2012} calculate the difference between the UV and optical bands, but $C$ is more sensitive to $\lambda_{\rm rf}$ at the short wavelength end.

The $A$ values also vary with the $\lambda_{\rm rf}$. We notice that the $A$ values all ``drop'' suddenly at the longest $\lambda_{\rm rf}$ in all redshift bins. \cite{Bignone2020} have pointed out that the $A$ value may be dominated by the $A_{bgr}$ when the $\langle$S/N$\rangle$ is low. The excessive correction can result in artificially low or even negative asymmetry values. We have compared the $A_{bgr}$ of different bands, and find that the $A_{bgr}$ of F444W corresponding to the longest $\lambda_{\rm rf}$, is systematically $\sim$0.02 higher than that of other bands indeed, most likely due to the relatively lower $\langle$S/N$\rangle$ of this band than other bands. This may result in an underestimation of $A$ at the F444W. This trend is consistent with the previous quantitative studies on the robustness of morphological measurement \citep[e.g., ][]{Lotz2004,Thorp2021,Treu2023}. If we ignore F444W bands, for each redshift bin we examined, ($z$=0.8-1.3, $z$=1.3-2.0, $z$=2.0-3.0), the $A$ values decrease by $\sim$0.022 ($\sim$13.3\%), $\sim$0.009 ($\sim$5.8\%), and $\sim$0.014 ($\sim$8.4\%) respectively, from rest-frame $\sim$570nm to $\sim$1000nm.

The $G$ parameter is similar to $C$, in the sense that higher $G$ means more concentrated light distribution. Therefore, $G$ is expected to have a similar wavelength-dependent trend with $C$. For each redshift bin, the $G$ values only increase by $\sim$0.016 ($\sim$3.3\%), $\sim$0.016 ($\sim$3.2\%), and $\sim$0.014 ($\sim$2.7\%), respectively, from rest-frame $\sim$570nm to $\sim$1340nm. The $M_{20}$, which is also an indicator of concentration, shows an anti-correlation with $\lambda_{\rm rf}$. For each redshift bin, the $M_{20}$ values decrease by $\sim$0.072 ($\sim$4.4\%), $\sim$0.085 ($\sim$5.3\%), and $\sim$0.034 ($\sim$2.2\%), respectively, from rest-frame $\sim$570nm to $\sim$1340nm. The trends of $G$ and $M_{20}$ are similar to \cite{Wuyts2012}, but the amplitude is lower. The reason is the same as $C$. The wavelength dependence of $G$ and $M_{20}$ at high redshift is also weaker than low redshift.

\begin{figure*}[ht!]
    \centering
    \includegraphics[width=1\textwidth]{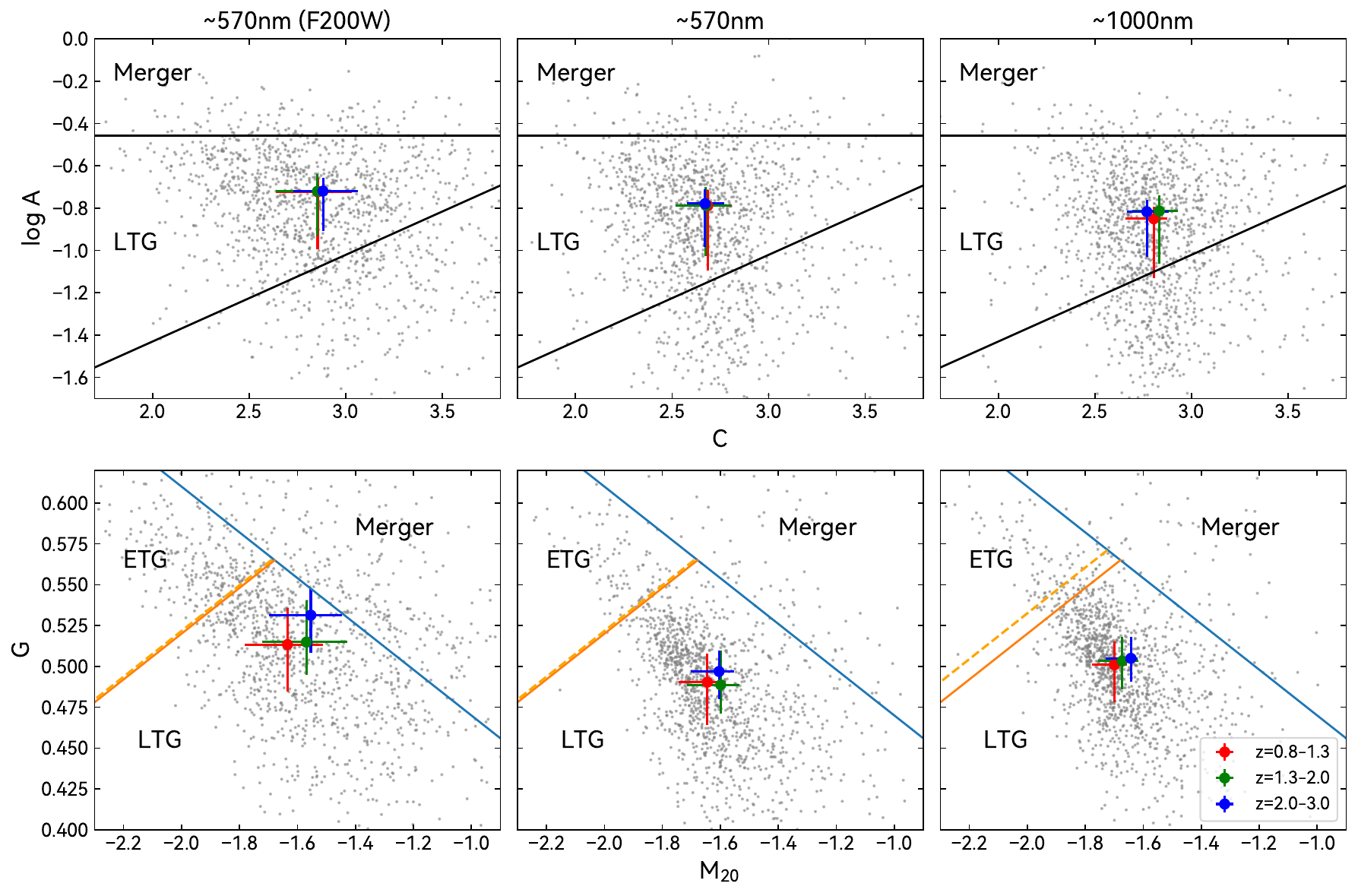}
    \caption{The $C-A$ and $G-M_{20}$ diagrams at rest-frame $\sim$570nm and $\sim$1000nm. The first column represents the values that have been matched to F200W PSF, and other columns represent the values that have been matched to F444W PSF. The red, green, and blue crosses are present the mean values and 1$\sigma$ quantiles of $z$=0.8-1.3, $z$=1.3-2.0, and $z$=2.0-3.0 bins, individually. The sloped lines on the $C-A$ diagrams are the boundaries between LTG and intermediate galaxies defined by \cite{Bershady2000}, and the horizontal lines are used to divide mergers and LTG at $A$=0.35 defined by \cite{Conselice2003}. The classification of galaxies on the $G-M_{20}$ diagrams is based on \cite{Lotz2008}. The solid orange lines ($G$=0.14$M_{20}$+0.80) show the boundaries between LTG and ETG, and the blue lines ($G$=$-$0.14$M_{20}$+0.33) divide mergers and non-mergers. The dashed orange lines are the boundaries between LTG and ETG described by Equation \ref{eq:corrected_etg_ltg}, which have corrected the $\lambda_{\rm rf}$-dependent morphology biases.
    \label{fig:g_m20_simplified}}
\end{figure*}

There is currently ongoing debate about the evolution of morphological parameters at $\lambda_{\rm rf}$. Some studies at $z<$6 have reported similar trends to our results \citep[e.g., ][]{Conselice2008,Wuyts2012}, while others at higher redshifts ($z>$6) have found no obvious wavelength dependence \cite[e.g., ][]{Yang2022,Treu2023}. \cite{Treu2023} suggests that the lack of dependence may be due to the shorter time available for star formation in high-redshift galaxies. The star formation of extremely high redshift galaxies selected by Lyman-break seems to be global. The old and young stellar populations are mixed together spatially, and cannot be separated by any filters. In contrast, galaxies at lower redshifts especially in local universe have more complex star formation histories, resulting in differences in the spatial distributions of old and young stellar populations. Our findings of the weaker wavelength dependence in the high-redshift bins are in line with this trend.

\subsection{$C-A$ and $G-M_{20}$ Diagrams} \label{subsec:c_a_g_m20}

Combinations of the morphological parameters can be used to classify galaxies into different types. The $C-A$ \citep{Bershady2000,Conselice2003} and $G-M_{20}$ \citep{Lotz2004,Lotz2008} diagrams are the most common non-parametric methods for classification. In this section, we utilize these diagrams to examine how the redshift, the $\lambda_{\rm rf}$, and the spatial resolution impact the distribution of galaxies in morphological parameter space.

\subsubsection{$C-A$ Diagram} \label{subsec:c_a}

The first row of Figure \ref{fig:g_m20_simplified} shows the $C-A$ diagrams of our galaxy sample at different $\lambda_{\rm rf}$ in different resolution. The diagram features the boundary that divides LTG and intermediate galaxy as proposed by \cite{Bershady2000} along with the threshold separating mergers from normal galaxies at $A$=0.35, defined by \citep{Conselice2003}. From these diagrams we observe that the average position of the galaxies (colored crosses) changes minimally with redshift but is affected by the $\lambda_{\rm rf}$ and resolution. When the $\lambda_{\rm rf}$ increases, C increases while A decreases, resulting in the galaxies shifting towards the lower right of the $C-A$ diagram. When the spatial resolution decreases, both $C$ and $A$ decrease, causing the galaxies to move in the lower left direction of the $C-A$ diagram. The impact of PSF will be discussed in more details in Section \ref{subsec:psf_influence}.



\subsubsection{$G-M_{20}$ Diagram} \label{subsubsec:g_m20}

\cite{Lotz2008} have defined a simple partitioning of the $G-M_{20}$ diagram that can be used to roughly separate ETG, LTG, and merging galaxies at $z$=0.2-0.4. The second row of Figure \ref{fig:g_m20_simplified} shows the $G-M_{20}$ diagrams of our galaxy sample at different $\lambda_{\rm rf}$ in different resolution. We can see that, unlike \cite{Lotz2008}, our sample does not exhibit a clear bimodality between LTG and ETG, which is similar to what has been seen at other redshifts and in simulations \citep[e.g.,][]{Lotz2008a,Bignone2020,Kartaltepe2023}. Instead, most of our galaxies are located in the LTG area in any $\lambda_{\rm rf}$, just like what is classified by the $C-A$ diagram in the first row of Figure \ref{fig:g_m20_simplified}. The 7th-9th columns of Table \ref{tab:bins_summary} list the quantity and fraction of each classification of the $G-M_{20}$ diagram.

The relationship between redshift and $G-M_{20}$ classification is also examined in conjunction with the results presented in Figure \ref{fig:wavelength_evolution}. As redshift increases, $M_{20}$ and $G$ exhibit an upward trend, leading to a shift of galaxies towards the upper right region of the $G-M_{20}$ diagram. Additionally, the fraction of mergers also increases. We note that the fraction of ETG decreases with the increase of redshift in all resolutions and $\lambda_{\rm rf}$, but the fraction of LTG decreases significantly in high spatial resolution (F200W) but increases in low resolution (F444W) in contradictory. This can be attributed to the large number of LTGs being reclassified as mergers in high resolution. This will be discussed more in Section \ref{subsec:psf_influence}.

\begin{figure}[ht!]
    \centering
    \includegraphics[width=1\columnwidth]{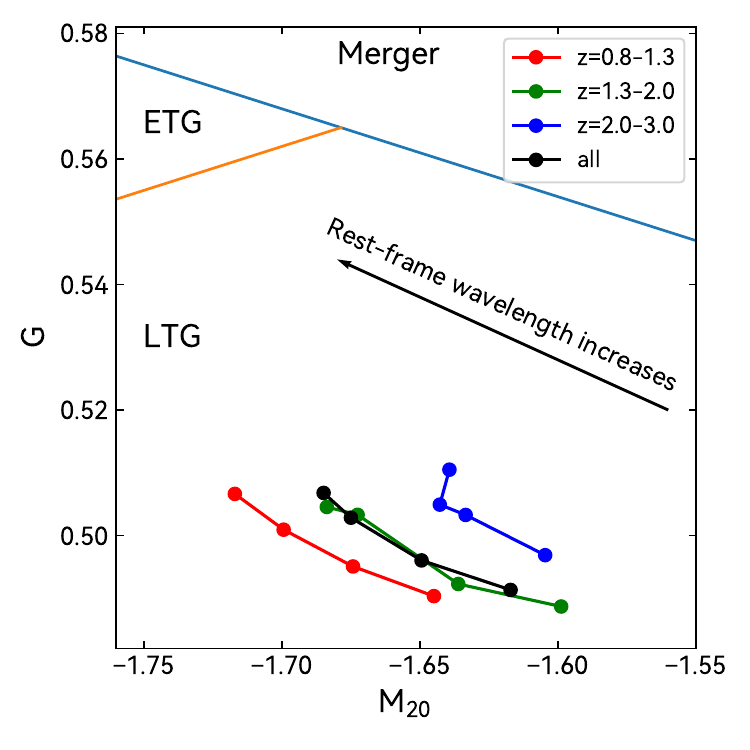}
    \caption{The movements of the mean value of galaxy morphological parameters on the $G-M_{20}$ diagram with the increase of $\lambda_{\rm rf}$. Different colors represent different redshift bins, and black dots represent the overall sample.
    \label{fig:g_m20_move}}
\end{figure}

Furthermore, the $\lambda_{\rm rf}$ is an additional factor that influences galaxy distribution in the $G-M_{20}$ diagram. Figure \ref{fig:wavelength_evolution} and \ref{fig:g_m20_move} show that an increase in $\lambda_{\rm rf}$ results in a decrease in $M_{20}$ and an increase in $G$, causing galaxies to move towards the upper left region of the $G-M_{20}$ diagram with a slope of -0.23$\pm$0.03. The path of the movement is almost parallel to the boundary between merger and non-merger, with an angle of less than 5$\arcdeg$. In terms of the proportion of different types shown in Table \ref{tab:bins_summary}, the fraction of LTGs decreases while the fraction of ETGs increases, with little change observed in the fraction of mergers. This reflects that the long rest-frame wave is more able to track the distribution of the old stellar population.

Because the division of \cite{Lotz2008} is based on the rest-frame $B$ ($\sim$450nm). To reduce potential wavelength-dependent biases in the morphological classification of galaxies, we suggest that the boundaries between LTG and ETG (the orange line in Figure \ref{fig:g_m20_simplified}) should rise with the increase of $\lambda_{\rm rf}$ to do the ``K-correction''. A simple modified form is
\begin{equation}
G = 0.14M_{20}+0.8+\alpha\left(\frac{\lambda_{\rm rf}~\rm nm}{450~\rm nm}-1\right). \label{eq:corrected_etg_ltg}
\end{equation}
The last item is the ``K-correction'' item and $\alpha$ is the correction factor. If the classification of a galaxy is not affected by $\lambda_{\rm rf}$, the ratio of LTG to ETG of the total sample should be stable at all $\lambda_{\rm rf}$. We obtain a correction factor of $\alpha=0.006$, which minimize the relative standard deviation of the ratios of LTG to ETG at different wavelengths for all redshift bins of our sample. When using the new corrected boundary, the ratio of LTG to ETG for our all galaxies is about 13.8:1. The ratio of our three redshift bins are about 11.0:1, 13.4:1, and 24.6:1, respectively. The corrected boundaries are shown as dashed orange lines in Figure \ref{fig:g_m20_simplified}. It should be noted that the above corrections are all carried out in the resolution of F444W (FWHM$\simeq$1.1kpc physically). This correction may not be applicable to other redshift ranges due to the potential different wavelength dependence of morphological parameters at different redshift. Another additional note is that the wavelength dependence of morphological parameters is not linear for an expanded wavelength range \citep[see Figure 3 in][]{Baes2020}, meaning that the corrected form of Equation \ref{eq:corrected_etg_ltg} may not be applicable to a wide wavelength range.

\subsection{PSF Influences on Morphological Measurements} \label{subsec:psf_influence}

In this section, we aim to investigate the impact of spatial resolution on the non-parametric morphology of galaxies. We not only calculate the morphological parameters convolved to F444W, but also calculate the parameters that have not been convolved and only convolved to F200W. The thin lines in Figure \ref{fig:wavelength_evolution} show the effect of different spatial resolutions on morphological parameters, and the first two rows of each bin in Table \ref{tab:bins_summary} show how the spatial resolution affects the classification of galaxies. We find that the smoothing effect of PSF can reduce $C$, $A$, and $G$ significantly ($\ga$10\%). The classification of galaxies is obviously affected by the size of PSF. The merger fraction in F200W resolution can be $\ga$2 times larger than that in F444W resolution, regardless of the classification method of $C-A$ or $G-M_{20}$. This is because if the spatial resolution decrease (i.e. the PSF FWHM increases), $C$, $A$, $M_{20}$, and $G$ would both decrease, the galaxies would move to the lower left direction in the $C-A$ and $G-M_{20}$ diagram. So, the fraction of merger would decrease significantly. This result can also explain why the line divided merger and non-merger galaxies defined by \cite{Lotz2008} is moved down compared with \cite{Lotz2004}. Because the sample of \cite{Lotz2004} is closer than \cite{Lotz2008}, and has a better spatial resolution. In addition, at the same time, the fraction of LTG has also increased with the decrement of resolution, because many of the merging galaxies in F200W resolution are classified as LTG in F444W resolution. 

\cite{Bignone2020} have used simulated images from the EAGLE simulation \citep{Schaye2015,Trayford2017} to study the effect of spatial resolution on non-parametric morphology. They have got the same trend as us, but they have found that $C$ only shows a small systematic deviation of less than 2\% in even for the worst FWHM of 1.5 kpc, which is much smaller than the variation we find. The average value of $C$ at $z$=2.0-3.0 in F115W resolution (FWHM$\simeq$0.33kpc) and F115W band (i.e. $\lambda_{\rm rf}\simeq$340nm) is $\sim$3\% and $\sim$10\% higher than that of F200W (FWHM$\simeq$0.54kpc) and F444W (FWHM$\simeq$1.18kpc) resolution respectively. This may be due to the differences in galaxy sample, where our sample at higher redshift consists of more irregular and clumpier galaxies that are more affected by PSF than their sample at $z$=0.1 which contains more elliptical ETGs that are less affected. This morphology-dependent statistics can also be found in our sample. For each redshift bin we examined ($z$=0.8-1.3, $z$=1.3-2.0, $z$=2.0-3.0), the $C$ values at rest-frame $\sim$570nm increase by $\sim$6\%, $\sim$7\%, and $\sim$8\%, respectively, from the PSF of F444W to F200W. Therefore, it is particularly important to perform PSF matching for morphological measurements of high redshift galaxies.

\subsection{Uncertainties of Morphological Measurements} \label{subsec:uncertainty}

It is difficult to obtain the uncertainties of morphological parameters directly, so we adopt the empirical relationship between $\langle$S/N$\rangle$ and error ($\Delta$) to estimate the uncertainties of our morphological parameters. \cite{Lotz2004} have found that $G$, $M_{20}$, and $C$ are reliable to within $\sim$10\% ($\Delta\leq$ 0.05, 0.2, and 0.3, respectively) for galaxy images with $\langle$S/N$\rangle\geq$2, and $A$ shows offsets less than 0.1 at $\langle$S/N$\rangle\geq$5.

During our sample selection process (Section \ref{subsec:sample}), we only include sources with $\langle$S/N$\rangle\geq$2 in all bands we analyzed, so the values of $\Delta$ given by \cite{Lotz2004} above can be considered as the upper limits of uncertainties. However, the average $\langle$S/N$\rangle$ of our all galaxies is $\ga$10 in all six bands, indicating that the actual typical uncertainties should be much smaller than the upper limits. Additionally, increasing the $\langle$S/N$\rangle$ cutoff to 5 during the sample selection process would not significantly affect the results (seeing Appendix \ref{subsec:increase_snr}). We also conduct a simulation moving low-redshift galaxies to higher redshift in Appendix \ref{subsec:redshift_move}. Therefore, we can conclude that the wavelength evolution we observed above in Figure \ref{fig:wavelength_evolution} and \ref{fig:g_m20_move} is significantly existed.

\subsection{Morphology Evolution for Different $M_*$} \label{subsec:stellar_mass}

\begin{figure*}[ht!]
    \centering
    \includegraphics[width=1\textwidth]{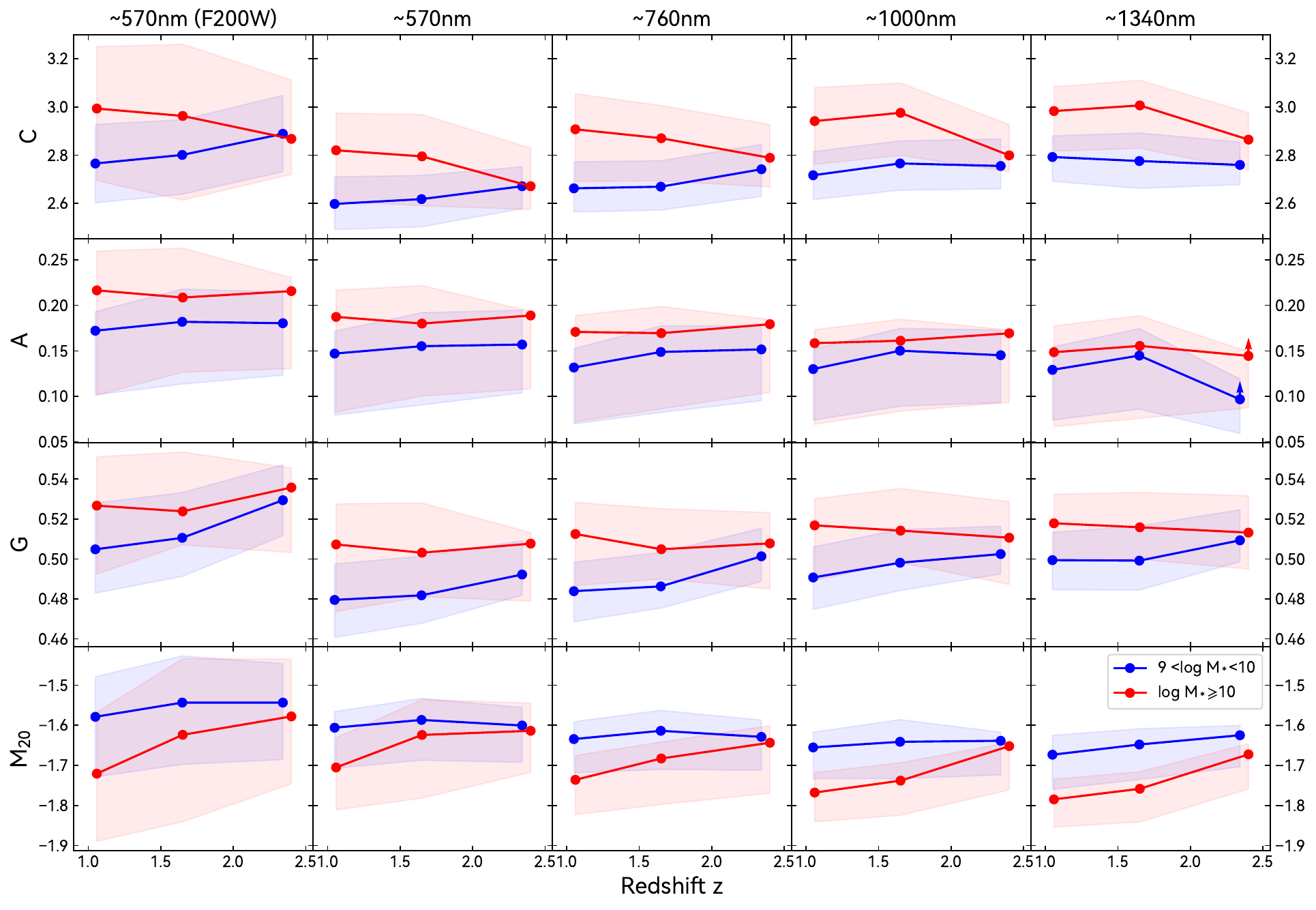}
    \caption{The four morphological parameters, $C$, $A$, $G$, and $M_{20}$ evolve with redshift at four matched $\lambda_{\rm rf}$ for the two stellar mass bins, 9$<$logM$_*<$10 and logM$_*\geqslant$10, colored by blue and red individually. The lines and dots represent the mean value in each bin. The semi-transparent filled areas indicate their 32\% and 68\% (1$\sigma$) quantiles. The first column represents the values that have been matched to F200W PSF, and the following columns represent the values that have been matched to F444W PSF. The arrows at the longest wavelength diagrams are the rough corrections of the underestimated $A$ in F444W.
    \label{fig:para_z}}
\end{figure*}

Galaxies with different $M_*$ usually have different evolutionary histories \citep{Madau2014}. We divide our sample into two $M_*$ bins (9$<$log$M_*<$10 and log$M_*\geqslant$10) to investigate the difference in morphological evolution between high-mass and low-mass galaxies. Figure \ref{fig:para_z} shows how the four morphological parameters, $C$, $A$, $G$, and $M_{20}$ evolve with redshift for different $M_*$. We can see that at $z\simeq$0.8-3.0, galaxies with log$M_*\geqslant$10 have higher $C$, $A$, and $G$, and lower $M_{20}$ than galaxies with log$M_*<$10 at any redshift bins, $\lambda_{\rm rf}$, and spatial resolutions. These differences are more significant at lower redshift bins, signifying different evolutionary paths for high- and low-mass galaxies. This evolutionary trend could not be caused by incompleteness of $M_*$ in the highest redshift bin, because the low-mass galaxies own lower values of $C$ and $G$, and higher values of $M_{20}$, while we find that the high-redshift bin which lacks low-mass galaxies has already owned these features. If we make the completeness of $M_*$ higher, this trend will only become stronger, rather than weaken.

In particular, for the evolution of $C$, our results are different from \cite{Whitney2021}. They have found that both high-mass and low-mass bins show a decrease in $C$ with cosmic time. This difference may be due to different observation bands and different data processing procedures. We discover that in blue $\lambda_{\rm rf}$ bands ($\lambda_{\rm rf}<$1000nm), the $C$ of galaxies with log$M_*\geqslant$10 increases with cosmic time, while the $C$ of galaxies with log$M_*<$10 decreases with cosmic time. In red rest-frame bands ($\lambda_{\rm rf}\geqslant$1000nm), the $C$ of both high-mass and low-mass galaxies increases with cosmic time. The $G$ of low-mass galaxies also demonstrates an evolutionary difference, albeit less pronounced than $C$'s. This evolutionary differences at different $\lambda_{\rm rf}$ can be explained as a scenario that the low-mass galaxies is still undergoing continuous star formation in the outside of their bulges, while the high-mass galaxies have already been assembled, and are experiencing merging mainly during this period \citep{Conselice2014,Madau2014} with higher $A$. The inside-out formation process leads to different evolutionary trends in different bands.

\section{Summary and Conclusions} \label{sec:conclusion}
Benefiting from the multi-wavelength broadband images from JWST/NIRCam, we have made use of the data processed by {\tt \string grizli} to investigate the non-parametric morphology of galaxies across the redshift ranges of $z\simeq$0.8-3.0 in the CEERS field. Our analysis have taken into account the PSF difference among different wave bands, and we have measured $C$, $A$, $G$, and $M_{20}$ of a total of 1376 galaxies divided into three redshift bins. The main results are as follows:

\begin{enumerate}
    \item Non-parametric morphological parameters change with the $\lambda_{\rm rf}$ at $z\simeq$0.8-3.0, especially at the short wavelength end. Different parameters and different redshift bins have different sensitivities on $\lambda_{\rm rf}$. The wavelength dependence of high redshift is weaker than that at low redshift. $C$ and $A$ are more sensitive than $G$ and $M_{20}$.
    
    \item The wavelength dependence of the morphological parameters may affect their usage in the classification of galaxy types, especially the distinction between ETG and LTG. We suggest that the boundaries between LTG and ETG in $G-M_{20}$ diagram should shift upwards as the rest-wavelength increases. We give a form of correction (Equation \ref{eq:corrected_etg_ltg}, $\alpha$=0.006) that can keep the ratio of LTG to ETG stable at different $\lambda_{\rm rf}$ for our current sample.
    
    \item Spatial resolution also significantly affects the morphological parameters and classification of galaxies, especially the fraction of merger. The merger fraction in F200W resolution can be $\ga$2 times larger than that in F444W resolution, regardless of the diagnostic parameter combinations we adopt. The influences of PSF become stronger as the redshift increases.
    
    \item The morphological parameters (especially the concentration parameter $C$) of high-mass (log$M_*\geqslant$10) and low-mass (9$<$log$M_*<$10) galaxies have a different evolution with redshift. At high redshift ($z\gtrsim$2), the difference between high-mass and low-mass galaxies is small, while at low redshift ($z\lesssim$1.5), the difference appears. In short $\lambda_{\rm rf}$ bands ($\lambda_{\rm rf}<$1000nm), the $C$ of high-mass and low-mass galaxies have opposite evolutionary trend, while in longer rest-frame bands ($\lambda_{\rm rf}\geqslant$1000nm), the trend becomes similar for both high- and low-mass galaxies. This suggests different evolution processes between high-mass and low-mass galaxies. The star formation of low-mass galaxies still continues in their disk, while galaxy mergers have played an important role in the evolution of high-mass galaxies during the period of $z\simeq$0.8-3.0.
\end{enumerate}


\begin{acknowledgments}
This work is supported by the Strategic Priority Research Program of Chinese Academy of Sciences (Grant No. XDB 41000000), the National Science Foundation of China (NSFC, Grant No. 12233008, 11973038, 11973039), the China Manned Space Project (No. CMS-CSST-2021-A07), the Cyrus Chun Ying Tang Foundations and the Frontier Scientific Research Program of Deep Space Exploration Laboratory.
\end{acknowledgments}

%

\vspace{5mm}
\facilities{JWST}


\software{astropy \citep{Collaboration2022},  
          Numpy \citep{Harris2020},
          Matplotlib \citep{Hunter2007},  
          statmorph \citep{RodriguezGomez2019}, 
          reproject \citep{Robitaille2020},
          photutils \citep{Bradley2022},
          Cython \citep{Behnel2011}
          }



\appendix
\section{Performance evaluation of starmorph\_csst} \label{sec:performance}

\begin{figure}[ht!]
    \centering
    \includegraphics[width=0.4\columnwidth]{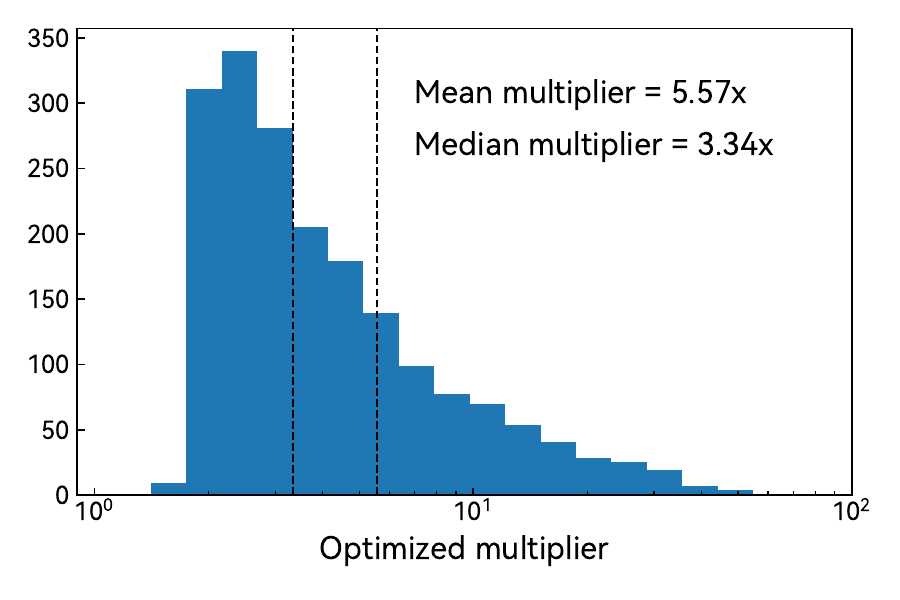}
    \caption{The distribution of the optimized multipliers, which is the ratio of the time consumed by the original version ({\tt\string statmorph}) to the time consumed by the optimized version ({\tt\string statmorph\_csst}).
    \label{fig:optimized_multiplier}}
\end{figure}

We make use of COSMOS/HST image to evaluate the multiplier of performance improvement achieved by {\tt\string statmorph\_csst}. We measured the CAS and $G-M_{20}$ morphological parameters of galaxies in the F814W band 064 image using both {\tt\string statmorph} and {\tt\string statmorph\_csst}, respectively, and record the time consumed for each galaxy. Finally, we get the morphological parameters of 1889 galaxies without any error flags. Figure \ref{fig:optimized_multiplier} shows the distribution histogram of the multipliers of performance, with an average optimization multiplier of $\sim$5.57, and a median of $\sim$3.34.

\begin{figure}[ht!]
    \includegraphics[width=1\columnwidth]{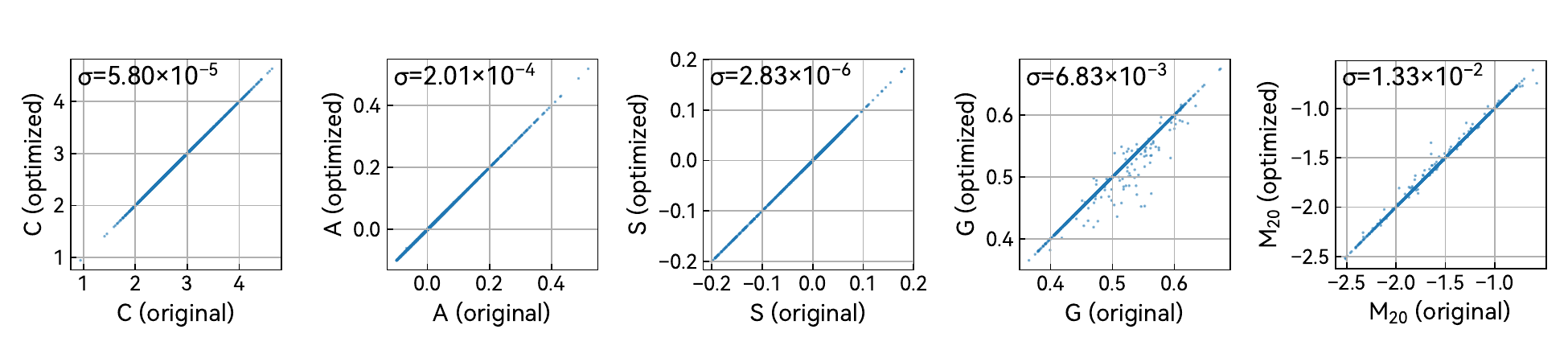}
    \caption{The comparison of the results of five main morphological parameters between {\tt\string statmorph\_csst} and the original {\tt\string statmorph}. $\sigma$ on the upper left is the standard deviation.
    \label{fig:cosmos_test}}
\end{figure}

Although we use Cython to rewrite the code of statmorph for optimization, considering that we reduce the max times of iteration during solving equations, and Cython and Python have different precision in calculating floating numbers, it is also necessary to verify whether the results of the optimized version differ significantly from the original version. Here we use the same sample above in the COSMOS field for testing. The results shown in Figure \ref{fig:cosmos_test} indicate that our optimization has almost no effect on the calculation of morphological parameters. Additionally, some other optimizations are also included in {\tt\string statmorph\_csst}, but it will not be enabled by default due to their significant increase in $\sigma$.

\section{The Influence of $\langle$S/N$\rangle$ on the Robustness of Measurement} \label{sec:noise_influence}

We employ two methods to assess the extent to which noise introduces significant biases in our morphological measurements. These methods include increasing the cutoff of $\langle$S/N$\rangle$ and conducting simulations for observations at higher redshifts.

\subsection{Increasing the Cutoff of $\langle$S/N$\rangle$} \label{subsec:increase_snr}

\begin{figure*}[ht!]
    \centering
    \includegraphics[width=0.85\textwidth]{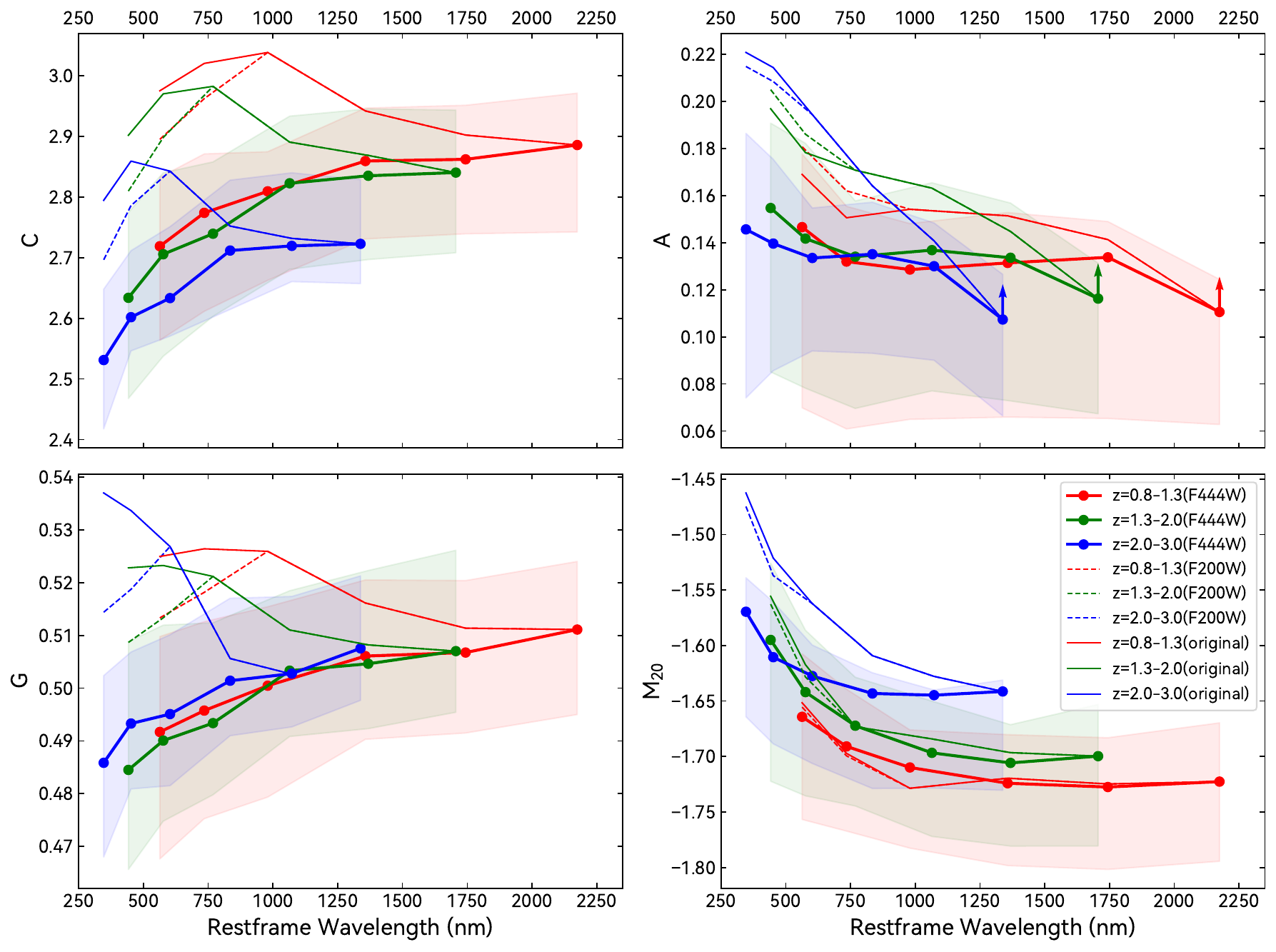}
    \caption{Same as Figure \ref{fig:wavelength_evolution}, but the cutoff of $\langle$S/N$\rangle$ increases to 5.
    \label{fig:wavelength_evolution_5}}
\end{figure*}

To reduce the uncertainties caused by noise and evaluate the influence of uncertainties on morphological parameter measurement. Here we increase the cutoff of $\langle$S/N$\rangle$ increases to 5 during the sample selection (Section \ref{subsec:sample}). The sample size is reduced to 770, and there are 378, 273, and 119 galaxies in each redshift bin. Figure \ref{fig:wavelength_evolution} shows how the average of the four morphological parameters vary with $\lambda_{\rm rf}$. Due to the discarding of low $\langle$S/N$\rangle$ sources, the underestimation of $A$ in F444W is weakened. So the arrows at the longest wavelength diagrams which present the rough corrections of the underestimation are shortened to 0.015. The wavelength evolution shown in Figure \ref{fig:wavelength_evolution_5} is not significantly different from Figure \ref{fig:wavelength_evolution}. 

\subsection{Moving the Low-redshift Images to Higher Redshift} \label{subsec:redshift_move}

In addition, We conduct simulations to move the images of galaxies from the low-redshift bin ($z$=0.8-1.3) to the mid-redshift ($z$=1.3-2.0) and high-redshift bin ($z$=2.0-3.0). The steps are as follows:
\begin{enumerate}

\item We cut out the F115W stamp images of the 504 galaxies in the low-redshift bin. The mid-redshift images will be observed in F150W, and the high-redshift images will be observed in F200W.

\item We resample the images and weight images to match the angular diameters at the target redshift. This relative variation of angular diameter is very weak, less than $\sim10\%$.

\item We darken the resampled images to match the surface brightness at the target redshift. The dimming coefficient is 
\begin{equation}
k=\left[\frac{D_A(z_{\rm target})}{D_L(z_{\rm target})} \times \frac{D_L(z_{\rm origin})}{D_A(z_{\rm origin})}\right]^2,
\end{equation}
where $D_L$ is the luminosity distance, $D_A$ is the angular diameter distance, $z_{\rm origin}$ is the original redshift before moving, and $z_{\rm target}$ is the target redshift. At high redshift, the cosmological dimming can have a significant impact on the surface brightness. For the high-redshift bin, the $k$ is about 0.1.

\item Due to the fact that the step above also weaken the noise of F115W images, It is necessary to add additional noise to the darkened images. The additional noise is generated from normal distribution whose $\sigma=\sqrt{\sigma_{\rm targetband}^2-k^2\sigma_{\rm F115W}^2}$. The $\sigma_{\rm band}$ is obtained from the weight images of bands. Thus, the simulated mid-redshift and high-redshift images from F115W will own the same noise level as the F150W and F200W respectively.

\item We convolve the original F115W images and the simulated higher-redshift images to the F444W PSF with the kernel created in Section \ref{subsec:psf_matching}, and re-measure their morphological parameters by {\tt \string statmorph\_csst}. We subtract the parameters before moving from the parameters after moving to obtain the $\Delta$ (i.e., $\Delta$=simulated $-$ original).

\end{enumerate}

\begin{figure*}[ht!]
    \centering
    \includegraphics[width=0.8\textwidth]{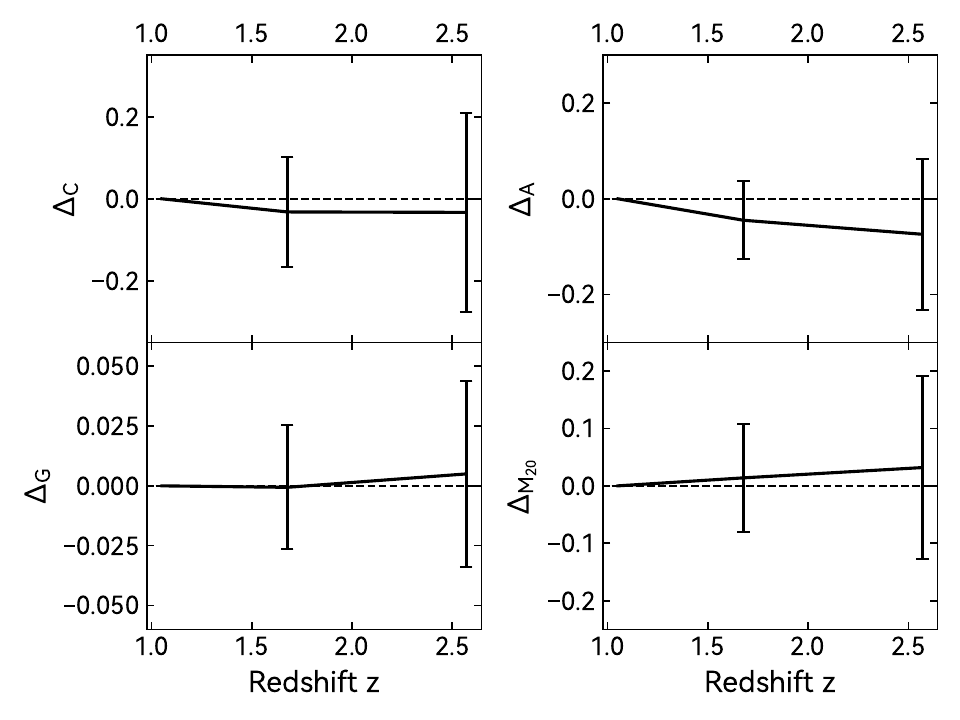}
    \caption{The systematic error (mean of $\Delta$) of $C$,$A$,$G$, and $M_{20}$ of the F115W images of galaxies in the low-redshift bin. The error bars are the random error (standard deviation of $\Delta$). Detailed data is listed in Table \ref{tab:redshift_move}.
    \label{fig:redshift_move}}
\end{figure*}

The results shown in Figure \ref{fig:redshift_move} and Table \ref{tab:redshift_move} 
 indicate that when moving to high redshift, the $\langle$S/N$\rangle$ deteriorates, and random errors increase significantly, but the systematic errors show minor changes. Since our results here represent the average values of numerous galaxies' parameters, the main consideration is the impact caused by systematic errors. Therefore, it can be proven that the evolution of morphological parameter with $\lambda_{\rm rf}$ and redshift is significant and reliable.

\begin{deluxetable}{ccccc}
	\tablecaption{Systematic and random error (mean and standard deviation of $\Delta$)\label{tab:redshift_move}}
	\tablewidth{0pt}
	\tablehead{
        \multirow{2}{*}{Parameter} & \multicolumn{2}{c}{Systematic error} & \multicolumn{2}{c}{Random error} \\
		 & \colhead{Move to F150W} & \colhead{Move to F200W} & \colhead{Move to F150W} & \colhead{Move to F200W}
	}
	\startdata
	$C$ & -0.0320 & -0.0331 & 0.1332 & 0.2425 \\
    $A$ & -0.0448 & -0.0741 & 0.0816 & 0.1575 \\
    $G$ & -0.0006 & 0.0050 & 0.0259 & 0.0388 \\
    $M_{20}$ & 0.0139 & 0.0318 & 0.0936 & 0.1591 \\
	\enddata
    \tablecomments{Move to F150W/F200W: move the low-redshift F115W images ($\lambda_{\rm rf}\sim570nm$) to the mid-redshift/high-redshift bin, which will be able to observed in the F150W/F200W filters. \\
    }
\end{deluxetable}


\bibliography{sample631}{}
\bibliographystyle{aasjournal}



\end{document}